\documentclass[sigconf]{acmart}
\usepackage{amsmath}
\usepackage{caption}
\usepackage{subcaption}
\usepackage{graphicx}
\usepackage{balance} 
\usepackage{vcell} 
\usepackage{array} 
\usepackage{multirow}
\AtBeginDocument{%
  \providecommand\BibTeX{{%
    \normalfont B\kern-0.5em{\scshape i\kern-0.25em b}\kern-0.8em\TeX}}}

\acmConference[WebSci'24]{16th ACM Web Science Conference 2024}{May 21--24, 2024}{Stuttgart, Germany}
\copyrightyear{2024}
\acmYear{2024}
\setcopyright{acmlicensed}\acmConference[Websci '24]{ACM Web Science Conference}{May 21--24, 2024}{Stuttgart, Germany}
\acmBooktitle{ACM Web Science Conference (Websci '24), May 21--24, 2024, Stuttgart, Germany}
\acmDOI{10.1145/3614419.3644000}
\acmISBN{979-8-4007-0334-8/24/05}

\begin{document}

\title{The role of interface design on prompt-mediated creativity in Generative AI}

\author{Maddalena Torricelli}
\affiliation{%
  \institution{City, University of London}
  \streetaddress{Northampton Square}
  \city{London}
  \country{United Kingdom}
  \postcode{EC1V 0HB}
}

\author{Mauro Martino}
\affiliation{%
  \institution{IBM Research}
  \streetaddress{75 Binney St}
  \city{Cambridge, MA}
  \country{United States}
  \postcode{02142}
}
\author{Andrea Baronchelli}
\email{andrea.baronchelli.1@city.ac.uk}
\affiliation{%
  \institution{City, University of London\\The Alan Turing Institute}
  \streetaddress{Northampton Square}
  \city{London}
  \country{United Kingdom}
  \postcode{EC1V 0HB}
}
\author{Luca Maria Aiello}
\email{luai@itu.dk}
\affiliation{%
  \institution{IT University of Copenhagen\\Pioneer Centre for AI}
  \streetaddress{Rued Langgaards Vej 7}
  \city{Copenhagen}
  \country{Denmark}
  \postcode{2300}
}
%

\renewcommand{\shortauthors}{}

\begin{abstract}
Generative AI for the creation of images is becoming a staple in the toolkit of digital artists and visual designers. The interaction with these systems is mediated by \emph{prompting}, a process in which users write a short text to describe the desired image's content and style. The study of prompts offers an unprecedented opportunity to gain insight into the process of human creativity. Yet, our understanding of how people use them remains limited. We analyze more than 145,000 prompts from the logs of two Generative AI platforms (Stable Diffusion and Pick-a-Pic) to shed light on how people \emph{explore} new concepts over time, and how their exploration might be influenced by different design choices in human-computer interfaces to Generative AI. We find that users exhibit a tendency towards exploration of new topics over exploitation of concepts visited previously. However, a comparative analysis of the two platforms, which differ both in scope and functionalities, reveals some stark differences. Features diverting user focus from prompting and providing instead shortcuts for quickly generating image variants are associated with a considerable reduction in both exploration of novel concepts and detail in the submitted prompts. These results carry direct implications for the design of human interfaces to Generative AI and raise new questions regarding how the process of prompting should be aided in ways that best support creativity.
\end{abstract}





\maketitle

\section{Introduction} \label{sec:intro}

Recent years have witnessed the rapid advancement of Generative AI~\cite{jovanovic2022generative}, a branch of artificial intelligence dedicated to developing computer models capable of creating original media such as text, images, speech, or music. Due to the impressive quality of their outputs, these models have drawn considerable public attention, which has been further sustained by the proliferation of increasingly accessible models~\cite{gozalo2023chatgpt} and openly available applications for generating a variety of synthetic content~\cite{brynjolfsson2023generative}. The rapid democratization of Generative AI~\cite{hu2023chatgpt} has arguably originated the largest shift in the participatory paradigm of the Web since the Web 2.0 revolution~\cite{anderson2007web}, with the crowd uploading massive amounts of AI-generated content online. This trend appears to be particularly pronounced in the realm of visual art, with an estimated 15 billions of AI-generated images uploaded to the Web as of August 2023~\cite{valyaeva23ai}. The collective process of mass creation of artistic content through artificial intelligence represents an unprecedented cultural phenomenon whose complex facets have attracted the interest of researchers from multiple disciplines~\cite{newton2023ai,de2023generative,oppenlaender2022creativity}.

One crucial aspect that characterizes the innovative creative process enabled by Generative AI tools lies in their user interface. The generation of visual artifacts is achieved through \emph{prompting}, a form of human-computer interaction that uses short descriptive texts as instructions for the AI models to create a desired image~\cite{oppenlaender2022prompt}. The ways in which people compose their prompts offers two significant insights into the behavioral aspects of AI-assisted creation of art. First, prompts embody the human creative intent, thereby serving as a valuable resource for understanding how people navigate the realm of ideas. Second, prompts are shaped around the distinctive capabilities and constraints inherent to the models' different interfaces, which presents an opportunity to quantify the extent to which design decisions regarding how Generative AI interacts with humans influence the creative process. Therefore, understanding the prompting practices that emerge during interactions with different AI tools is essential to uncover the behavioral mechanisms that define the cultural process of artistic production in the AI era and that are not yet fully understood.

In this context, we analyze textual prompts used for image generation to address two key questions:

\vspace{2pt}\noindent \emph{Q1)} To what extent people \emph{explore} different concepts in the process of creating generative art?

\vspace{2pt}\noindent \emph{Q2)} How do different designs of Generative AI tools either foster or hinder this exploration?

\vspace{2pt}By answering these questions, we aim to shed light on the process of human creativity itself, and offer insights on how technology can be designed to support it.

To answers these questions, we analyze more than 145,000 prompts from two text-to-image generator platforms which differ significantly in their structure, design, and objectives: DiffusionDB~\cite{wang2022diffusiondb} and Pick-a-Pic~\cite{kirstain2023pick}. We examine the sequences of prompts submitted by each user in terms of \emph{exploration} and \emph{exploitation}, two aspects of a decision making pattern that refers to how individuals balance between exploring new available options and exploiting familiar choices to optimize a given outcome~\cite{he2023modified,toyokawa2014human,baronchelli2013levy}.
In the context of this study, users can either engage in exploration by experimenting with unfamiliar prompts, or they can lean towards exploitation by composing prompts similar to those they have previously submitted, indicating a preference for specific image styles or content. We quantify these patterns by measuring the evolution of textual similarity between prompts and visual similarity between images within user sessions. In our analysis, we adopt the working hypothesis that exploration is loosely yet positively associated with creativity: given a fixed number of interactions with the platform, users who submit more diverse prompts create on average a wider range of artistic content.

Our findings reveal that users have a propensity towards exploration over exploitation. However, the presence of features to generate new image variants with a simple click, without revising previously submitted prompts, is associated both with a reduced degree of exploration and shorter prompts. Our findings raise new questions about how to best aid user interactions with Generative AI tools without constraining the creative process that arises from those interactions.

\section{Dataset} \label{sec:data}

\begin{figure}[t!]
\centering
\includegraphics[width=0.99\columnwidth]{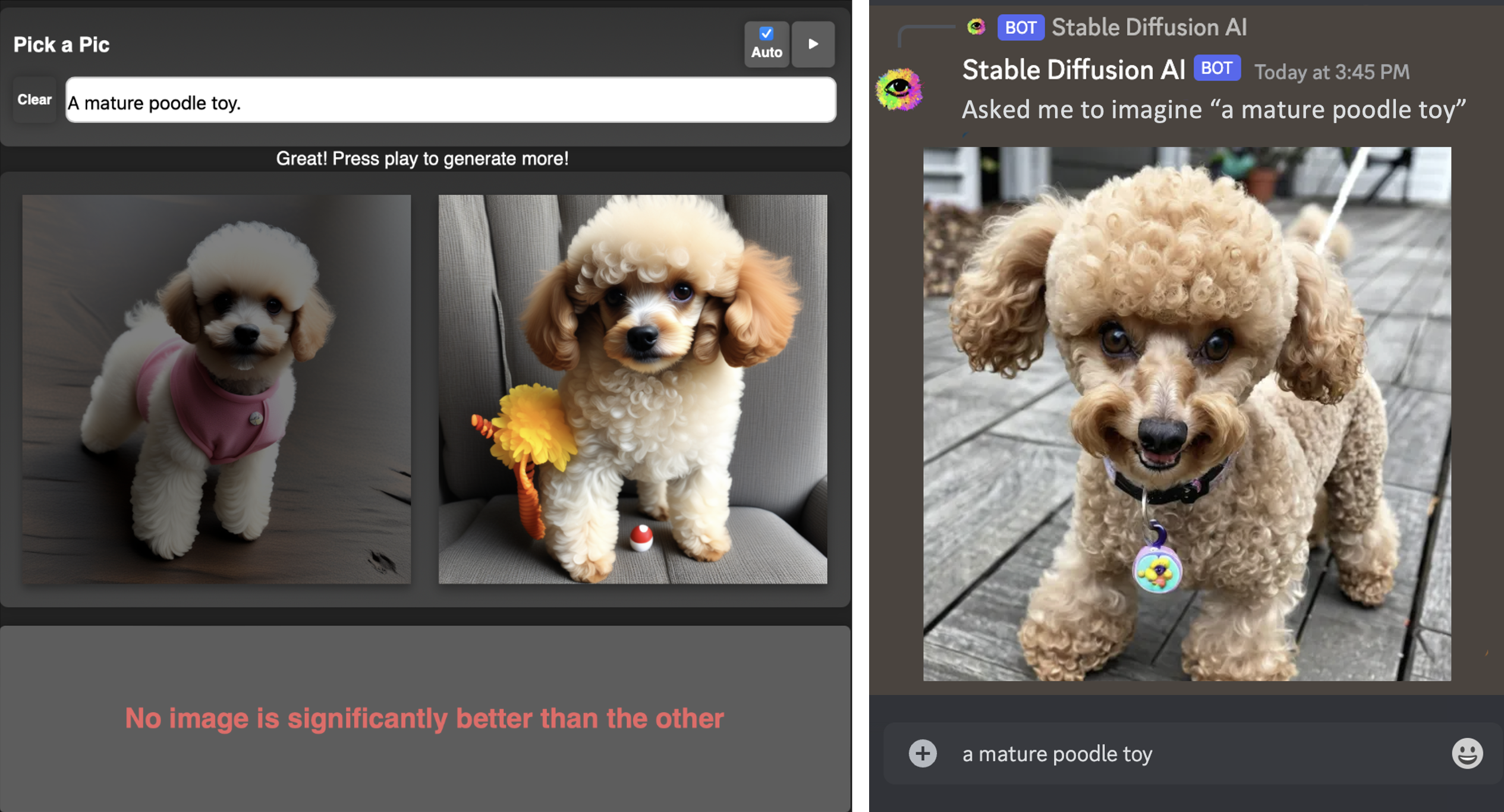}
\caption{User interface of Pick-a-Pic (left) and Stable Diffusion (right).}
\label{fig:platform_screenshots}
\end{figure}

Figure~\ref{fig:platform_screenshots} shows the interfaces of both platforms we collected data from. To interact with the Stable Diffusion model, users submit prompts via dedicated channels on the official Stable Diffusion Discord server and receive automated responses containing the corresponding generated image. DiffusionDB~\cite{wang2022diffusiondb} is a large-scale repository of images created by the Stable Diffusion model, paired with the prompts that users submitted to generate them. This data collection was obtained by scraping the Stable Diffusion Discord channel at the beginning of its activity, from 1st to 31st August 2022. The collection contains 10,177 unique users and 1.5 million prompts, each prompt paired with the image it generated.

Pick-a-Pic is a web platform that allows users to submit prompts for image generation and nudges them to provide annotations of image quality. Upon receiving a prompt, the platform generates a pair of image variants and invites the user to select their preferred option. The non-preferred variant is replaced with a new one derived from the same prompt, and the new pair can be rated again. At any time, users can decide to cast a vote with a simple click or to generate a different pair of images by writing a new prompt. In the dataset, a click on the interface counts as a submission of the same prompt it was used in the previous interaction. The platform creators logged all user interactions in January 2023 (the month in which the platform launched) and compiled them into a dataset comprising 992 users and 77,929 prompts.

It is important to stress that the data from both platforms was collected right after the platforms were launched, and a time in which very few open tools for AI image generation were accessible to the general public. It is therefore reasonable to assume that all users were roughly equally familiar with this type of platforms during the data collection period.

All images are free from illegal, hateful, or NSFW content, and are published under CC0 license that waives all copyright and allows uses for any purpose. The information from the official Stable Diffusion Discord server is public. The Pick-a-Pic dataset was obtained previous explicit consent from the platform users.

For each dataset, we focus on the essential data to study temporal patterns of human-AI interaction: the \emph{userID}, the textual \emph{prompt}, the \emph{timestamp} of prompt submission, and the set of \emph{images} returned. We call \emph{prompt sequence} a user's temporally-sorted list of prompts.

The platforms exhibit different levels of popularity. To mitigate potential activity biases when comparing user sessions across both platforms, we harmonize their characteristics by selecting a subset of users from DiffusionDB that align more closely with the activity of Pick-a-Pic users. To achieve that, we match each of the 992 Pick-a-Pic users with a randomly chosen DiffusionDB user who had submitted a number of prompts within 50\% of those submitted by the corresponding Pick-a-Pic user. To ensure the robustness of our results to this step of random selection, we repeat this sampling process ten times with replacement, and report all our results as the average over these ten randomized selections. On average, the DiffusionDB samples contain 70,081 prompts.

\section{Methods} \label{sec:methods}

\begin{figure}[t!]
\centering
\includegraphics[width=0.99\columnwidth]{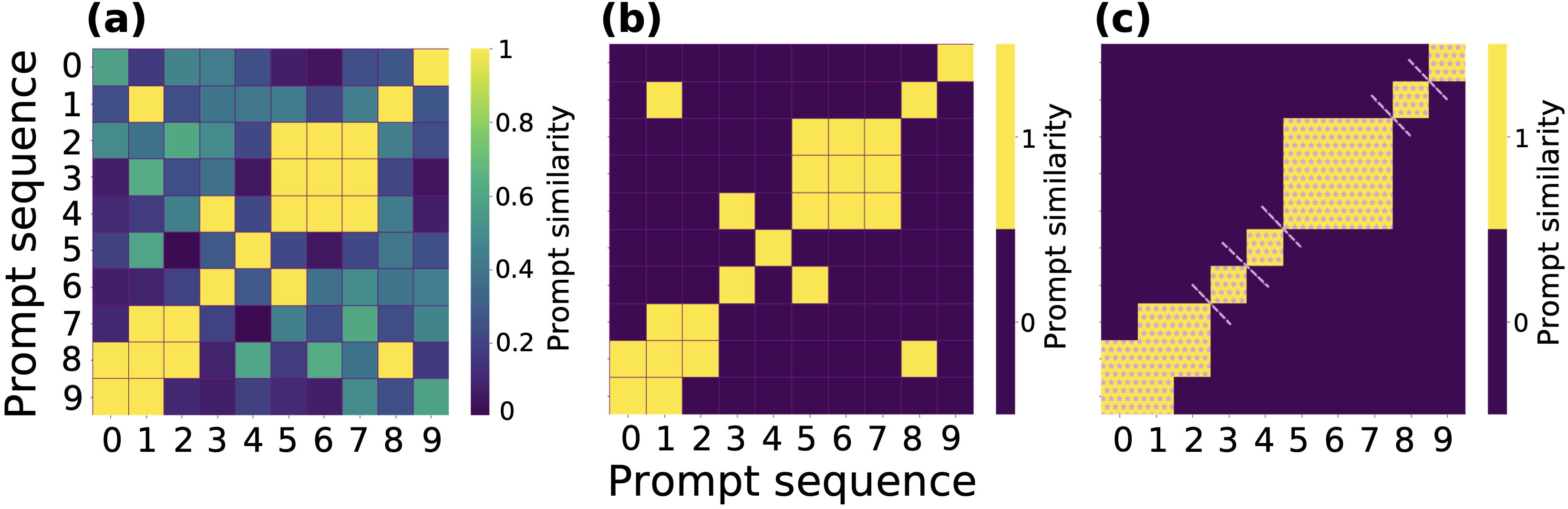}
\caption{To identify topical transitions in prompting, we calculate a similarity matrix between pairs of prompts sorted by their submission time (a), binarize the matrix (b), and identify blocks of highly-similar, consecutive prompts around the matrix diagonal (c).}
\label{fig:pipeline_jumps_jaccard_sim}
\end{figure}

\noindent \textbf{Image and prompt similarity.}
For each user, we calculate the similarity between all pairs of images they generated. To achieve that, we map each image into a numerical embedding space using DinoV2 \cite{oquab2023dinov2}, a family of foundation visual models designed to extract image features that are suitable for a number of image processing tasks, including calculating visual similarity. DinoV2 models achieve state-of-the art performance across most common benchmarks, outperforming other openly available weakly-supervised models, such as OpenCLIP~\cite{ilharco2021openclip}. Specifically, we used the \texttt{lvd142m} model, that produces embeddings with 769 dimensions. Given two DinoV2 vectors $\mathbf{v_{image_i}}$ and $\mathbf{v_{image_j}}$, we calculate the cosine similarity between them:
\begin{equation}
\text{sim}(image_i,image_j) = \frac{\mathbf{v_{image_{i}}} \cdot \mathbf{v_{image_{j}}}}{\|\mathbf{v_{image_{i}}}\| \|\mathbf{v_{image_{j}}}\|}.
\end{equation}
We also calculate the similarity between all pairs of prompts submitted by any given uses. We do that using the Jaccard index, namely the intersection over the union of the set of tokens composing prompts $i$ and $j$:
\begin{equation}
\text{sim}(prompt_i,prompt_j) = \frac{|tokens_i \cap tokens_j|}{|tokens_i \cup tokens_j|}.
\end{equation}

\noindent \textbf{Topical variation.}
The degree of exploration or exploitation that users exhibit during their interaction with the AI can be quantified by examining the pairwise similarity between their prompts or images. In particular, the similarity between consecutive prompts in a user's sequence being low indicates a \emph{topical variation} in that user's focus. This is illustrated in Figure~\ref{fig:pipeline_jumps_jaccard_sim}(a), which presents the Jaccard similarity matrix of the prompts of an example user. Here, rows and columns represent the user's prompt sequence, and the cells encode the Jaccard similarity between all prompt pairs in the sequence. High similarity values along the diagonal suggest an exploitation process, where the user submits a series of consecutive prompts that are semantically similar. Conversely, low similarity scores around the diagonal indicate exploration, where consecutive prompts are markedly different from each other.

To measure the frequency of topical variations in a user's prompts, we first binarize the similarity matrix. This is achieved by assigning a value of 1 to elements above the matrix average and 0 to the rest, as shown in Figure~\ref{fig:pipeline_jumps_jaccard_sim}(b). We then count the number of 1-valued cell \emph{blocks} adjacent to the diagonal (Figure \ref{fig:pipeline_jumps_jaccard_sim}(c)). A boundary between blocks is identified when the binarized similarity between two consecutive prompts is zero ($sim_{bin}(prompt_i, prompt_{i+1}) = 0$). Finally, we quantify the number of topical variations as the number of blocks minus one ($n_{variation} = n_{blocks} -1 $). This number ranges from one (when all prompts are similar) to the number of prompts minus one (when each prompt is dissimilar from all others). From this count, we calculate the probability of a topical variation for each user:

\begin{figure}[t!]
\centering
\setlength\tabcolsep{0pt}
{
    \begin{tabular}[t]{cc}
    \begin{tabular}[t]{c}
        \begin{subfigure}{0.47\columnwidth}
            \centering
            \includegraphics[width=0.99\textwidth]{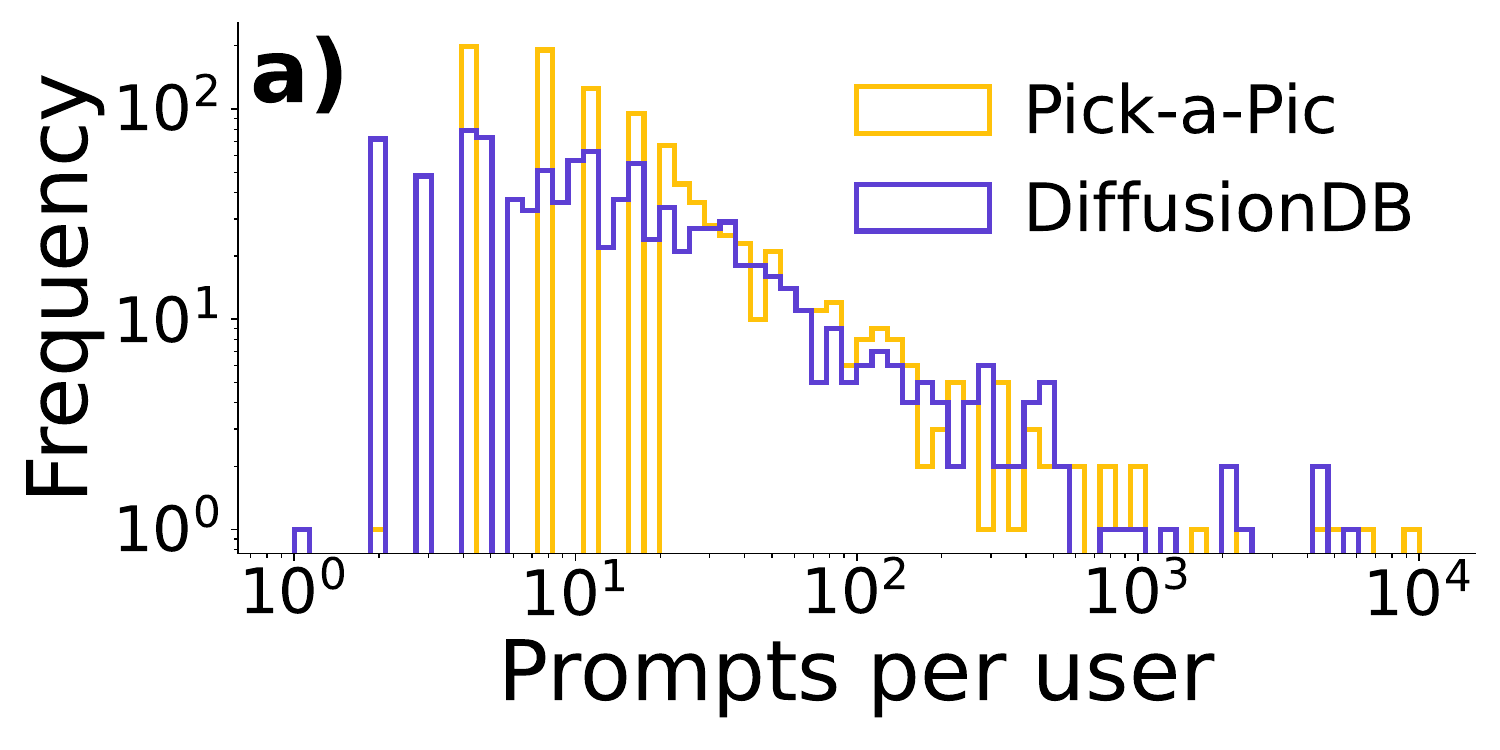}
        \end{subfigure} \\
        \begin{subfigure}{0.47\columnwidth}
            \centering
            \includegraphics[width=0.99\textwidth]{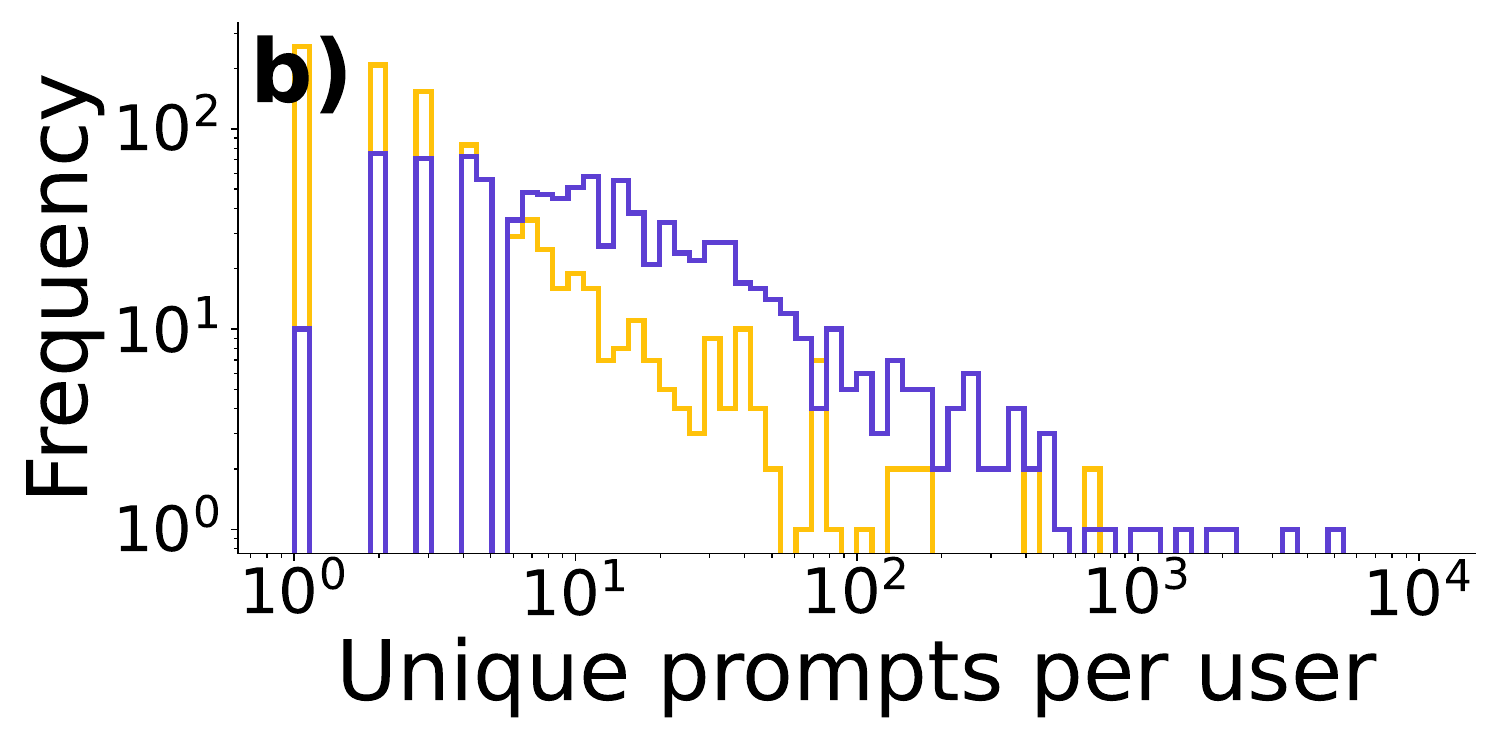}
        \end{subfigure}
    \end{tabular} &

    \multirow{1}{*}[50pt]{
    \begin{subfigure}{0.50\columnwidth}
      \includegraphics[width=0.99\textwidth]{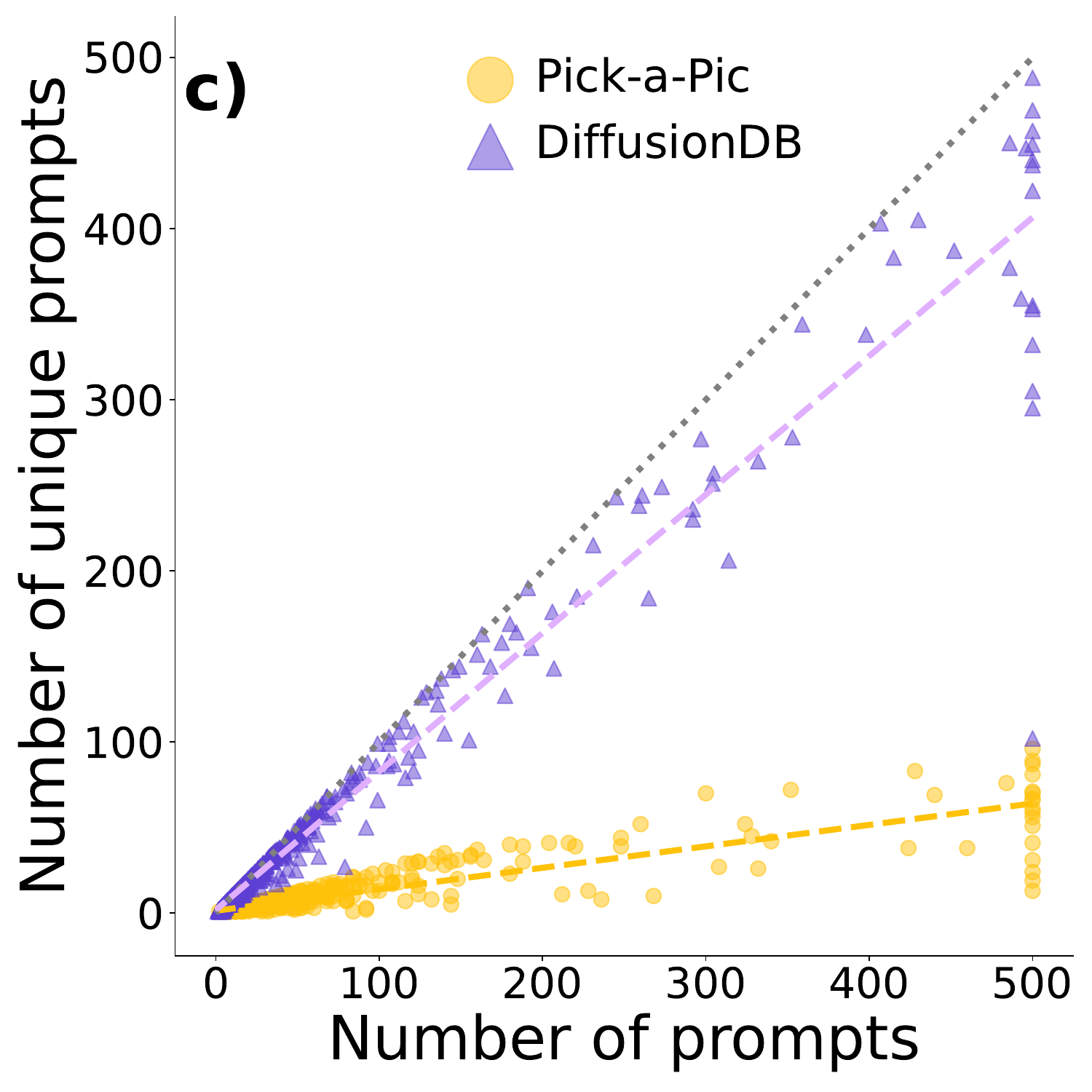}
    \end{subfigure}}
    \end{tabular}
}
\caption{Distribution of number of prompts (a) and unique prompts (b) per user. Total number of unique prompts submitted after $n$ interactions (c).}
\label{fig:stats}
\end{figure}

\begin{figure}[t!]
\centering
\includegraphics[width=0.99\columnwidth]{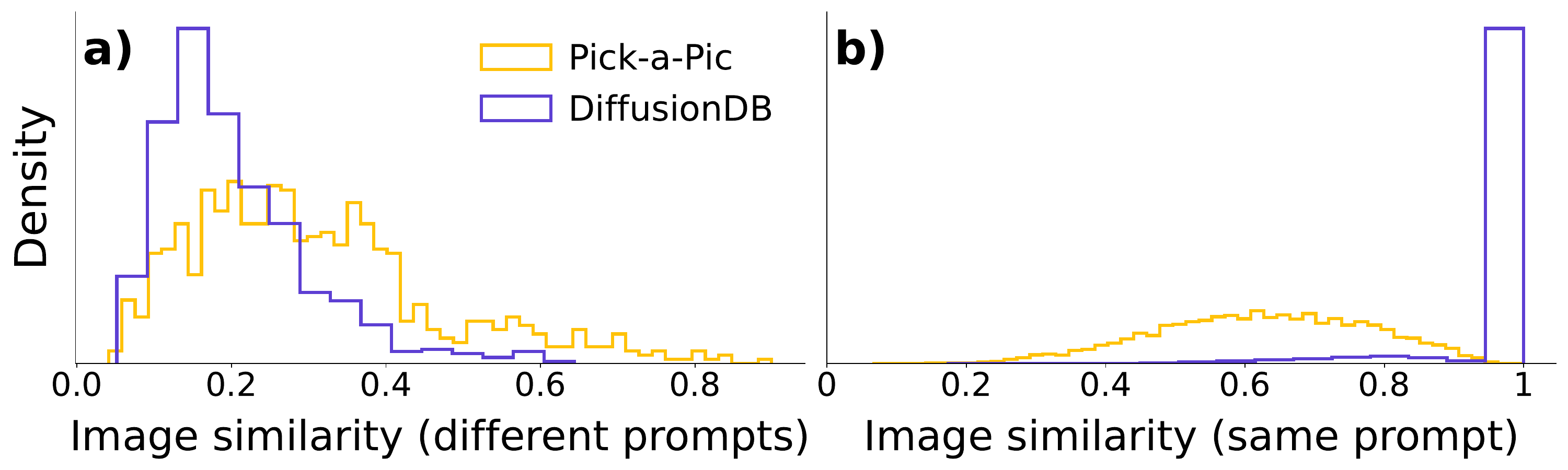}
\caption{Probability distributions of image similarity for: (a) all pairs of different prompts; (b) all pairs of identical prompts within the same prompt sequence.}
\label{fig:fig_platform_design}
\end{figure}
\begin{equation}\label{eq:prompt_variation_prob}
    P(\text{variation}) = \frac{n_{variation}}{n_{prompt}-1} = \frac{d_{blocks}-1}{n_{prompt}-1}.
\end{equation}

\section{Results} \label{sec:results}

\subsection{Image and prompt similarity} \label{sec:results:stats}

The distribution of the number of prompts per user is broad, with the majority of users submitting a limited number of prompts ($median=12$ in Pick-a-Pic, $median=11$ in DiffusionDB), while few power users submit thousands during the observed period (Figure~\ref{fig:stats}a). The data preprocessing we conducted to align the two datasets results in a comparable distribution of prompts per user across both platforms (two-sample Kolmogrov Smirnoff test $p=0.0$). However, the platforms' distinct interaction designs lead to significant differences in the types and range of prompts users submit and, consequently, the variety of images produced. Specifically, every time a user rates an image in Pick-a-Pic, the system generates a new image variant for the same prompt, while Stable Diffusion tends to produce identical or very similar variants for identical prompts (Figure~\ref{fig:fig_platform_design}).

\begin{figure}[t!]
\centering
\includegraphics[width=0.99\columnwidth]{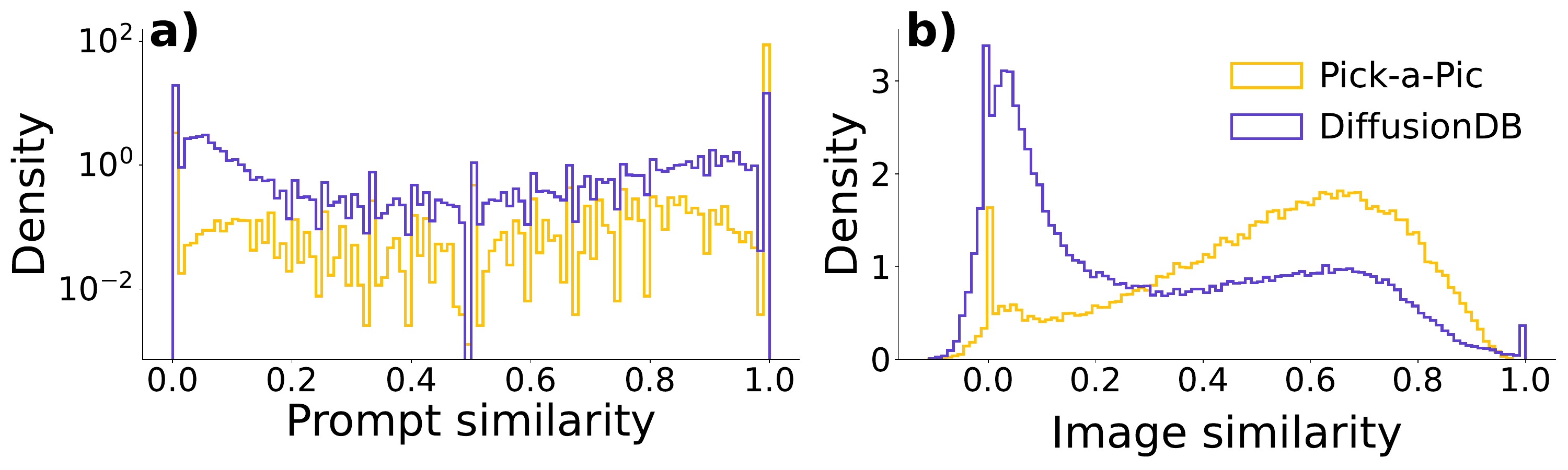}
%{figs/SI/extra/fig_diad_cosine_jaccard_sim.eps}
\caption{Distribution of similarity of consecutive prompts (a) and consecutive images (b) in a user sequence.}
\label{fig:fig_diad_cosine_jaccard_sim}
\end{figure}

\begin{figure}[t!]
    \begin{subfigure}{0.34\columnwidth}
      \includegraphics[width=0.99\textwidth]{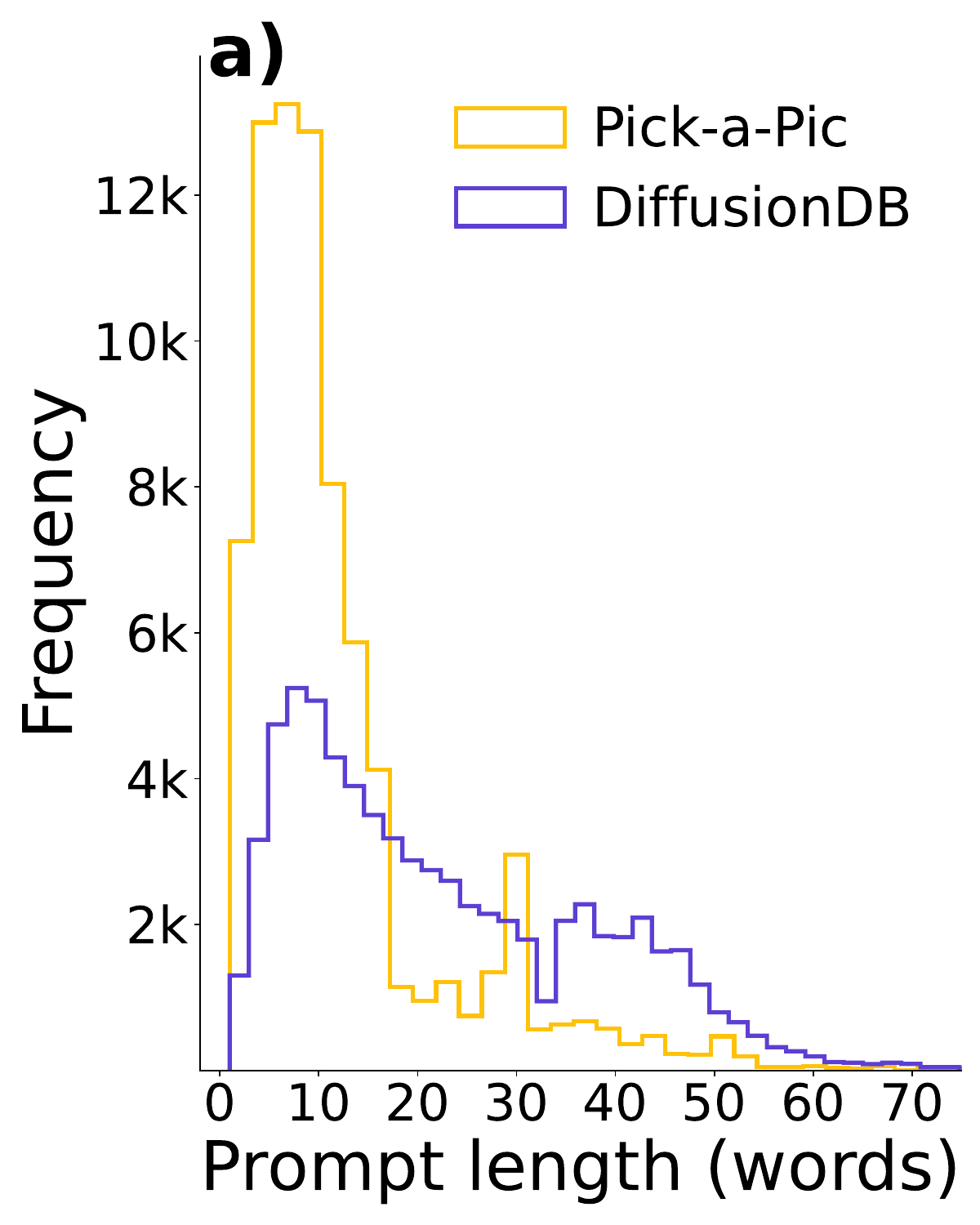}
    \end{subfigure}
    \begin{subfigure}{0.63\columnwidth}
      \includegraphics[width=0.99\textwidth]{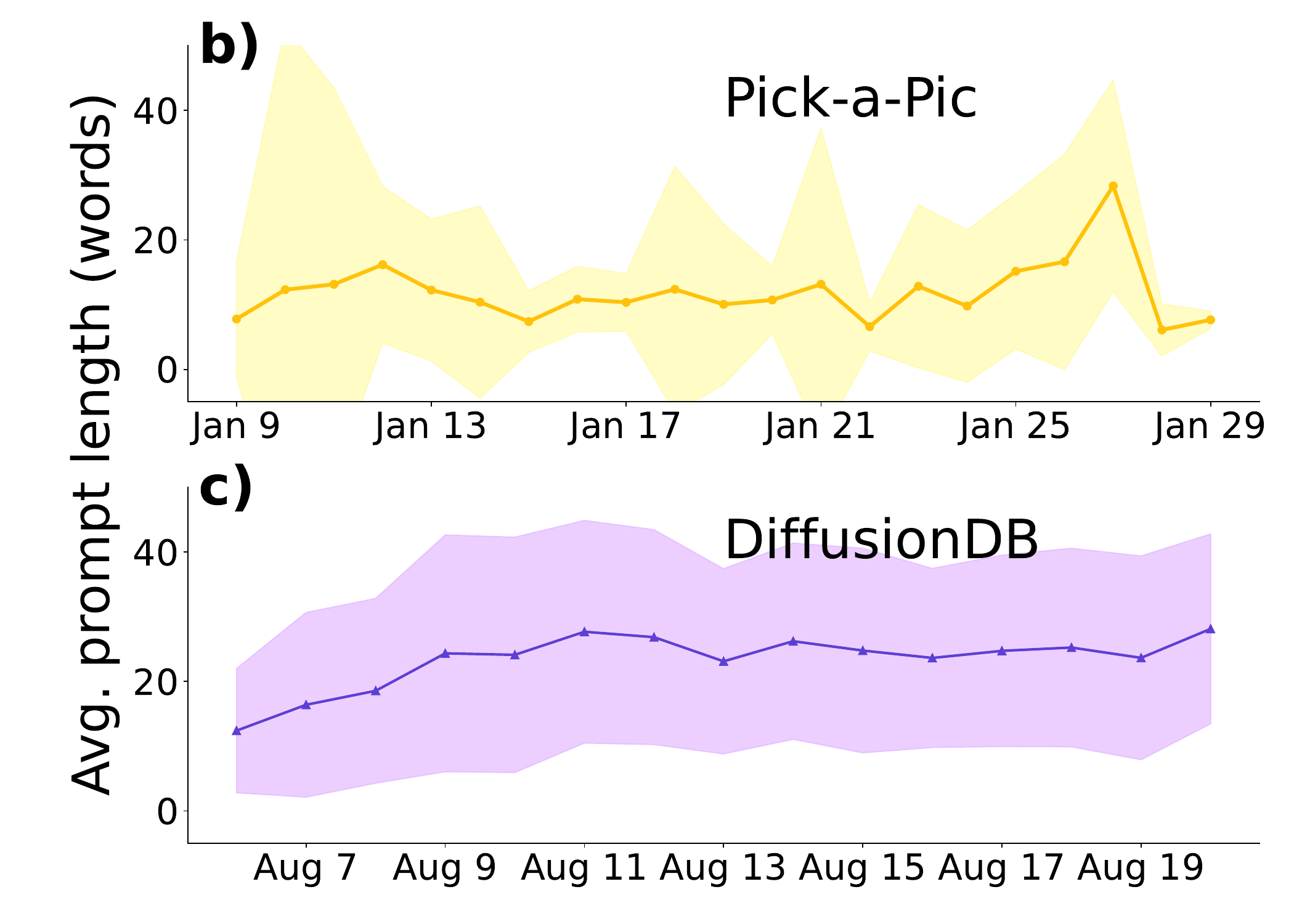}
    \end{subfigure}
\caption{Distribution of number of words per prompt (a) and evolution of average prompt length over time (b,c).}
\label{fig:fig_info_prompt_length}
\end{figure}

The average number of \emph{unique} prompts submitted by users is significantly lower in Pick-a-Pic than in Stable Diffusion ($avg.=8.7$ vs. $avg.=58.0$), Figure~\ref{fig:stats}b). Given the option to generate new images with a simple click without revising their prompt, Pick-a-Pic users typically explore fewer prompts than Stable Diffusion users after the same number of rounds, and this gap widens with the number of interactions with the AI (Figure~\ref{fig:stats}c). These distinct practices directly influence the similarity between consecutive prompts and images. In Pick-a-Pic, the distribution of similarity between consecutive prompts is relatively even, with the exception of notable peaks at the extreme values of zero and one, indicating users either sticking to the same prompt when generating the next image or switching to a completely non-overlapping prompt (Figure~\ref{fig:fig_diad_cosine_jaccard_sim}a). Conversely, the same similarity distribution in DiffusionDB shows a slight U-shape and a lower frequency of pairs with similarity equal to 1, suggesting that repetitions of the same prompts are less common, and there is a tendency towards more gradual prompt modifications. These two different approaches to prompting directly impact the variety of images generated. The distributions of visual similarity for consecutive images in a users' sequence are bimodal for both platforms, with two peaks around 0 and 0.7 indicating processes of exploring new visual content and exploiting variations of content generated with the previous prompt in the sequence (Figure~\ref{fig:fig_diad_cosine_jaccard_sim}b). However, DiffusionDB exhibits a more pronounced exploration peak, while Pick-a-Pic shows a more prominent exploitation peak.

These results indicate that an AI system capable of generating a diverse set of images from the same prompt may discourage users from trying new prompts. More surprisingly, Pick-a-pic prompts are not only repeated more frequently, but they are also significantly shorter (Figure~\ref{fig:fig_info_prompt_length}a). This suggests that interfaces facilitating easy ways to generate images without the need of prompting may divert user attention from the process of prompt composition, potentially resulting in missed opportunities for users to learn how to craft their prompts to enhance the quality of the desired output. This hypothesis is supported by the trend of the average length of prompts over time (Figure~\ref{fig:fig_info_prompt_length}b): the average prompt length remains roughly constant in Pick-a-pic, while it increases steadily in Stable Diffusion, a possible indication that Stable Diffusion users learn to refine and improve their prompts over time.

\subsection{Exploration vs. exploitation in prompting} \label{sec:results:exploration}

\begin{figure}[t!]
    \begin{subfigure}{0.34\columnwidth}
      \includegraphics[width=0.99\textwidth]{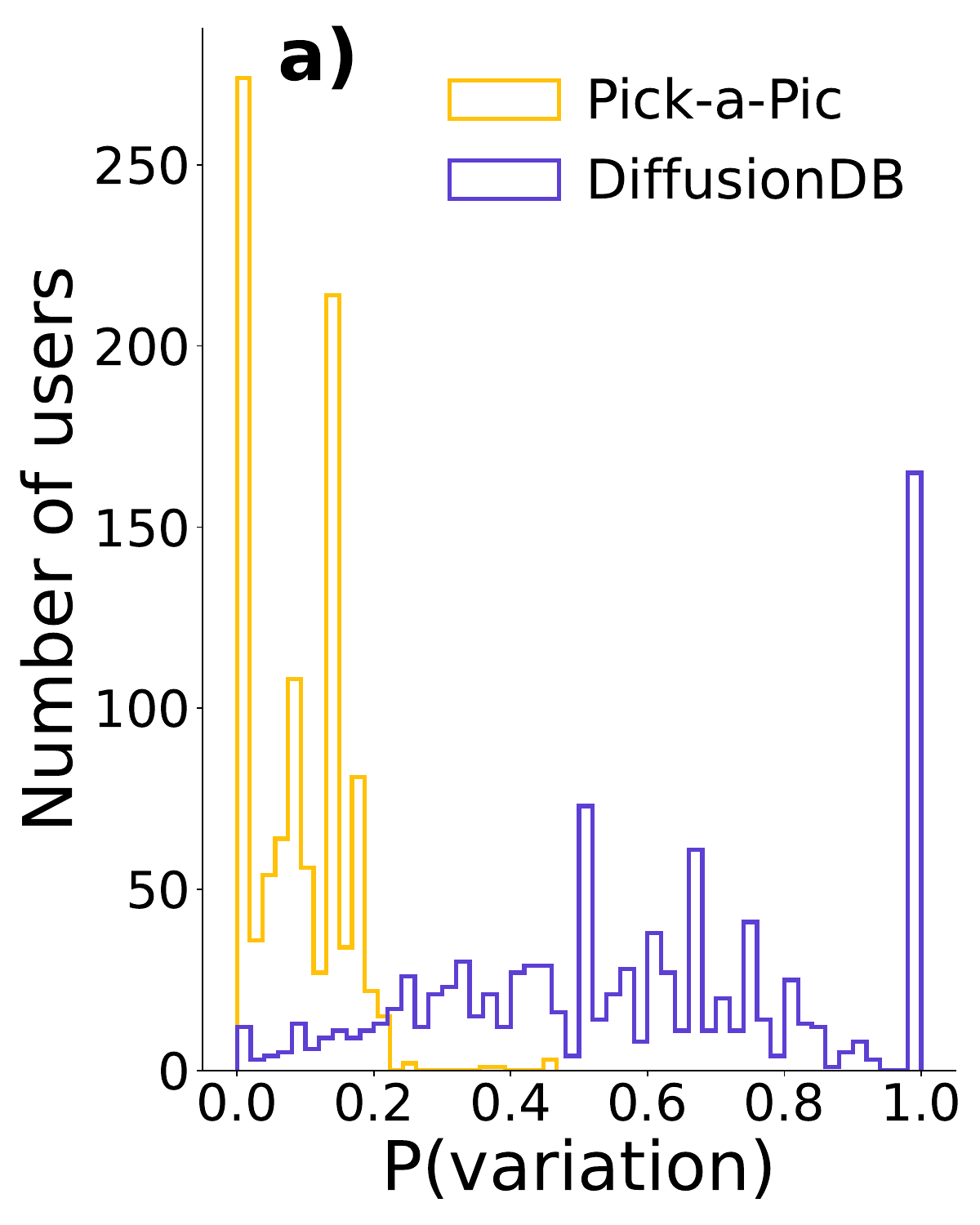}
    \end{subfigure}
    \begin{subfigure}{0.63\columnwidth}
      \includegraphics[width=0.99\textwidth]{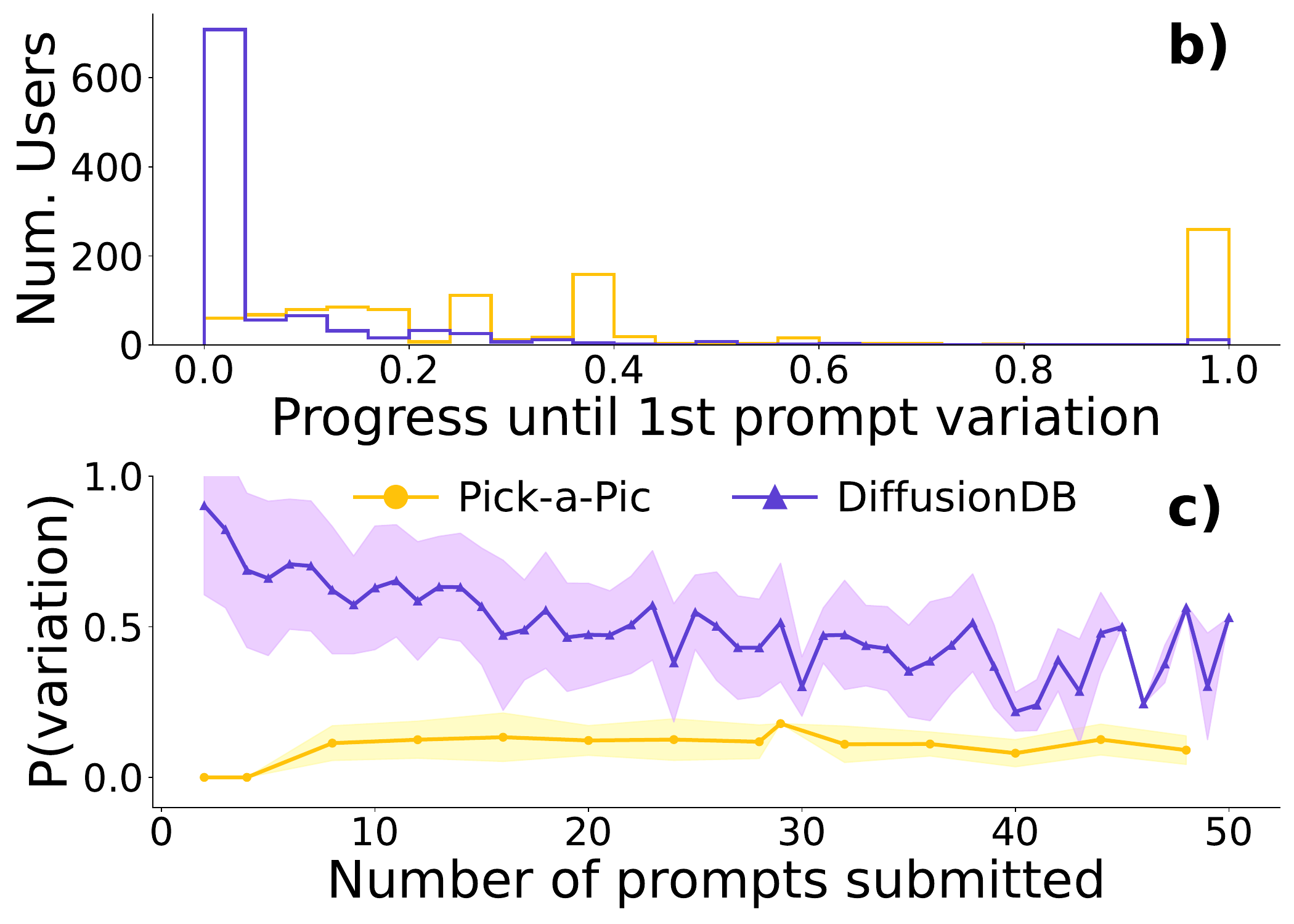}
    \end{subfigure}
\caption{Topical variation in prompting. (a) Distribution of the user-specific probability of changing topic when submitting a new prompt. (b) Number of users who experienced their first topical variation at a given point of their prompt sequence, normalized by the length of the sequence. (c) Average probability of topical variation for users who submitted a fixed number of prompts; standard deviation is shown.}
\label{fig:fig_triad_probability}
\end{figure}

The diversity of user-submitted prompts varies significantly across platforms, leading to two distinct experiences in terms of the breadth of visual content that users encounter during their interaction with the AI. To quantify these differences within the context of an exploration-exploitation process, we aggregate similar consecutive prompts for each user and calculate the probability $P(variation)$ of exploring an entirely new topic with their next prompt, as defined in Formula~\ref{eq:prompt_variation_prob}. The probability distributions reveal a stark contrast between the two platforms (Figure~\ref{fig:fig_triad_probability}a). Users of Stable Diffusion exhibit a wide range of behaviors, but the majority tend towards exploration, with $63\%$ changing topics more often than not ($P(variation) \geq 0.5$) and $16\%$ changing topic at every prompt ($P(variation) = 1$). Conversely, almost all Pick-a-Pic users change topic less than once every five prompts, and $26\%$ of them never do. Moreover, most Stable Diffusion users explore new topics immediately after their first prompt, whereas Pick-a-Pic users typically attempt topical variations much later in their sequence of interactions with the AI (Figure~\ref{fig:fig_triad_probability}b). The proportion of exploratory behavior over time changes significantly in Stable Diffusion, decreasing from approximately $0.9$ at the second prompt to around $0.5$ after $15$ prompts (Figure~\ref{fig:fig_triad_probability}c). This trend aligns with a scenario where users gain confidence in trying prompt variations within the same topic as they gain experience with the AI interaction. In contrast, the probability of a topic transition remains stable over time in Pick-a-Pic, suggesting that its human-computer interaction paradigm does not encourage users to alter their approach to prompt variation, at least within the timeframe of our dataset.

\begin{figure}[t!]
\centering
\includegraphics[width=0.99\columnwidth]
{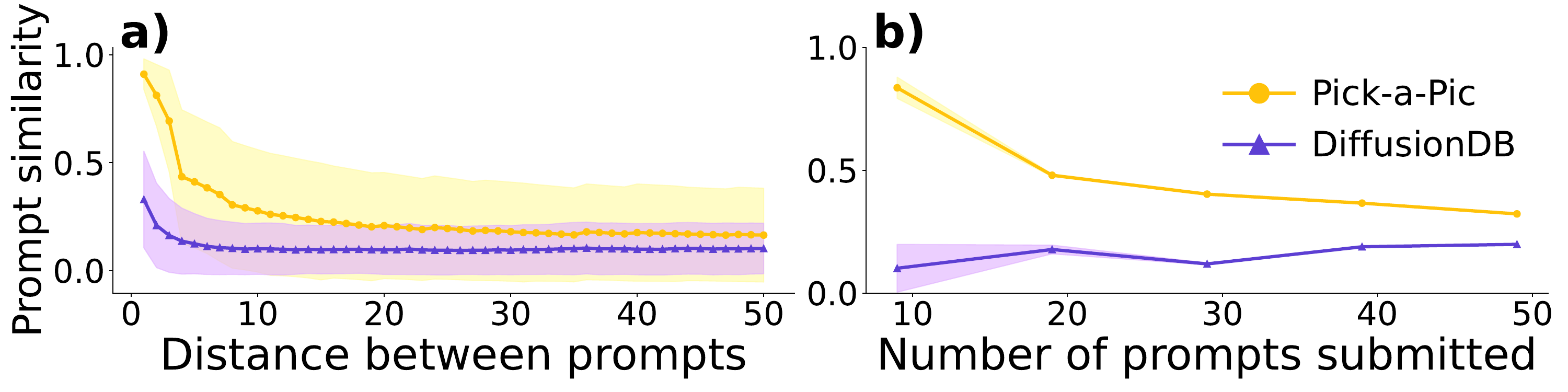}
\includegraphics[width=0.99\columnwidth]
{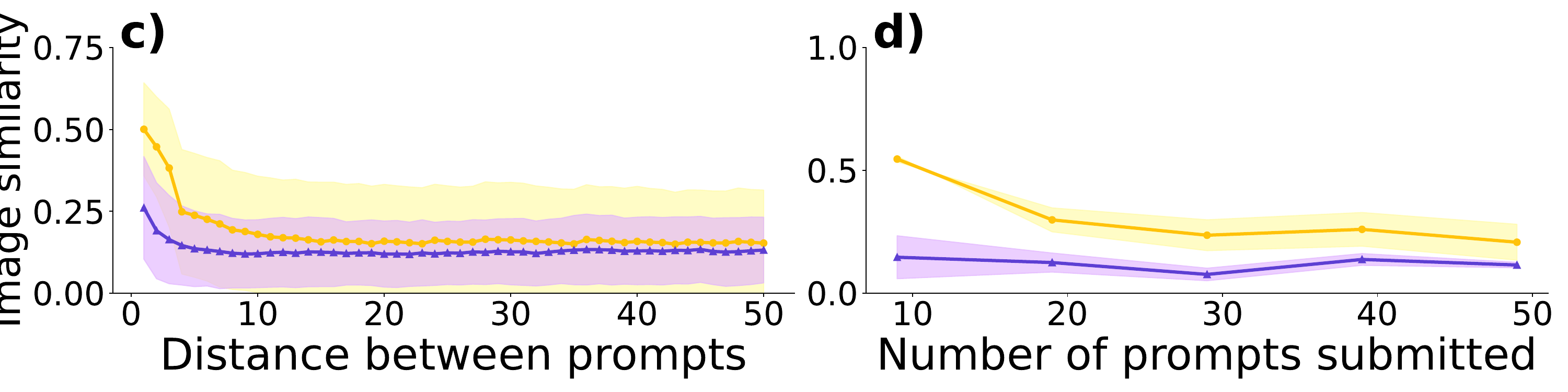}
\caption{Similarity of prompts and images over time. Average and standard deviation of: (a) the similarity between pairs of prompts submitted by the same user at distance $d$ in their sequence of prompts; (b) similarity between all pairs of prompts of users who submitted a fixed number of prompts. Panels (c) and (d) show the same trends for image similarity.}
\label{fig:fig_triad_avg_prompt_sim}
\end{figure}

The Pick-a-Pic feature that allows for generating new images without the need for prompt revision appears to slow down exploration compared to the Stable Diffusion interface that solely relies on prompting to interact with the AI interaction. This raises the question of how many prompts it takes for the average Pick-a-Pic users to explore a topical space as diverse as that typically navigated by Stable Diffusion users. To address this, we measure the similarity of prompt pairs at a distance $d$ within a user's prompt sequence to gauge the rate of topical drift (Figure~\ref{fig:fig_triad_avg_prompt_sim}a). Three trends become apparent. First, as anticipated by our prior analysis, the starting offsets of the similarity curves differ noticeably between the two platforms, with the similarity at $d=1$ in Pick-a-Pic being approximately three times greater than that in Stable Diffusion. Second, similarity in both platforms decreases rapidly with $d$. In Pick-a-Pic, the steepest rate of decay occurs at $d=4$, aligning with the observation that users typically do not transition to new topics until after five image generations. In Stable Diffusion, the average similarity stabilizes around $0.1$ after just a few iterations. Last, the similarity curve of Pick-a-Pic converges with that of Stable Diffusion at slow pace. At $d=4$, the standard deviation intervals of the two curves begin to overlap, and only at $d=8$ the average Pick-a-Pic similarity becomes comparable with the average in Stable Diffusion at $d=1$. This suggests that a Pick-a-Pic user requires approximately 4 to 8 iterations to experience a topical shift equivalent to that typically experienced by Stable Diffusion users in a single iteration. In aggregate, the average diversity of images explored by Pick-a-Pic users approaches, but never quite equals, that of Stable Diffusion users (Figure~\ref{fig:fig_triad_avg_prompt_sim}b). Similar patterns emerge when considering image similarity instead of prompt similarity (Figure~\ref{fig:fig_triad_avg_prompt_sim}c,d). In conclusion, while users across both platforms demonstrate a similar propensity for exploration, mechanisms that divert attention from prompt editing significantly impede this process.

\section{Conclusions} \label{sec:concl}

Generative AI solutions are quickly becoming a staple in the digital toolbox for computer-aided art production. Ensuring that these tools function as expected requires aligning their design with their intended use. We examined the prompt logs of two synthetic image generation platforms through the lens of the exploration-exploitation trade-off, and obtained two findings that have direct implications for the design of Generative AI interfaces.

\vspace{3pt}\noindent \emph{A1)} When their interaction with the AI is focused exclusively on prompting, users tend to explore a diverse range of prompts rather than reusing previously seen concepts, often shifting topics entirely from one prompt to the next. 

\vspace{3pt}\noindent \emph{A2)} When Generative AI tools provide functionalities for generating image variations without composing new prompts, the exploration process significantly slows down, prompts become less complex, and the breadth of content to which users are exposed is reduced.

\vspace{3pt}These findings suggests that while Generative AI systems hold a great potential for inspiring diversity in creative thinking, this potential may be curtailed by the introduction of supportive functionalities that limit prompt modification.

Our study is constrained by its focus on two platforms and the inability to control for external factors such as differing user attitudes towards prompting, art, or technology. These factors and other confounders that are not accounted for in this study might contribute to the observed differences between the two platforms. To fully account for these factors and confirm the generality of our findings, it is necessary to extend our study to a broader range of platforms. More in general, our study opens the question of how to strike a good balance between building accessible interfaces that alleviate the burden of composing complex prompts and creating the best conditions for users to express and enhance their creativity  while interacting with Generative AI tools.

\section{Related work} \label{sec:related}

Our work is inspired by previous research in this domain and builds upon it by providing a quantitative study of a pattern of human-computer interaction that has not been quantified so far. In the domain of image generation, prompting has been primarily explored from the user experience perspective through small-scale qualitative studies. Prior research has employed surveys and focus groups to delve into the motivations behind the use of AI art, indicating that a significant proportion of users are motivated purely by recreational use~\cite{sanchez2023examining}. A few interview-based studies have provided insights into the strategies people employ to approach prompting, examining aspects such as how people evaluate and enhance the quality of their prompts~\cite{oppenlaender2023prompting} the cognitive processes involved in composing prompts~\cite{schleith2022cognitive}, and the strategies for exploring prompt variations~\cite{zamfirescu2023johnny}. Only a few large-scale quantitative studies have been conducted to date. For example, Xie et al. compared a vast corpus of image prompts with traditional query logs from Web search engines, discovering that prompts are typically longer and semantically broader than traditional queries~\cite{xie2023prompt}. Vodrahalli et al. measured the diversity of chains of prompt reformulations from a dataset of 50k human-AI interactions, observing that diversity does not decrease as users refine their prompts~\cite{vodrahalli2023artwhisperer}. 

\bibliographystyle{ACM-Reference-Format}

\balance

\end{document}